# QUANTUM ROTORS AND THEIR SYMMETRIES*

Jerzy Dudek

Université Louis Pasteur, Strasbourg I *and* Institut de Recherches Subatomiques
F-67037 Strasbourg, France

AND

Andrzej Góźdź and Daniel Rosły

Institute of Physics, M. Curie–Skłodowska University
PL-20031 Lublin, Poland

A connection between nuclear symmetries other than those of an ellipsoidal nucleus and the properties of the implied rotational spectra are discussed. The discussion is focussed on a few examples of exotic shapes predicted recently by microscopic calculations. Some possible interpretation difficulties related to experiment are shortly mentioned.

## 1. Introduction

Quantum rotors have been extensively studied in the past and turned out to be powerful theoretical tools in exploring the *microscopic symmetries* of the examined objects, but the variety of possibilities that their theory offers has been explored mainly in molecular physics. Nuclear physics applications, although numerous, have been primarily limited to the quantum Hamiltonians corresponding to a classical rotating ellipsoid. Formally such Hamiltonians are invariant with respect to a four-element point group composed, in addition to the identity element, of three rotations through the angle of $\pi$ about the three principal axes of the reference frame. This group is denoted $D_2$; the corresponding rotors are said to be $D_2$-symmetric. An excellent presentation of this important particular case of a nuclear symmetry exists in Chapter 4 of Ref. [1], where numerous limiting cases and several approximate analytical expressions can be found.

Quantum rotor Hamiltonians are by definition operator functions of the angular momentum components $\{\hat{I}_\mu;\ \mu = \pm 1, 0\}$ only. No analog of the potentials as functions of the coordinates (here: rotation angles) exists for isolated nuclear or molecular rotors. Although common in various applications, the corresponding Hamiltonians are rather exotic objects. They can be viewed upon as composed exclusively of the 'kinetic energy' operator, the angular momentum playing a role analogous to that of the linear momentum in the usual kinetic energy expressions. Secondly, the parity, a concept so natural in quantum mechanics, can only be introduced with some special efforts, the angular momentum being a pseudo-vector rather than a vector.

Unlike the usual kinetic energy operators that are limited to the quadratic order expressions in $\{\hat{p}_x, \hat{p}_y, \hat{p}_z\}$, the rotor Hamiltonians are in general *not* limited to the second order expressions in $\{\hat{I}_\mu\}$. In molecular physics very successful applications exist for the Hamiltonians that are of sixth or higher orders. The presence of high-order terms expresses a non-rigid, many body structure of the corresponding quantum objects. Various symmetries present in molecules can be successfully modeled by the appropriately constructed Hamiltonians that are of order higher than 2 and may simulate any symmetry in question. We follow this line of thought here aiming at the nuclear physics applications.

In nuclear spectra encountered in experiments we are confronted with the overwhelming presence of the rotational bands, i.e. the sequences of energies $E_I$ that satisfy $E_I \sim I(I+1)$ and are often composed of very many transitions. In some cases these bands can be very well approximated by a simple parabolic rule $E_I = a\, I(I+1)$ with a single constant $a$, in other cases such an approximation is barely satisfactory, in many cases not satisfactory at all[1]. It will be one of our goals here to examine among others, the deviations of the curvature of the energy *vs.* spin relations from the simplest parabolic rule for (selected) symmetries of the rotors.

We are going to study the rotor Hamiltonians of the following general structure

$$\hat{\mathcal{H}} = \frac{\hat{I}_x^2}{2\mathcal{J}_x} + \frac{\hat{I}_y^2}{2\mathcal{J}_y} + \frac{\hat{I}_z^2}{2\mathcal{J}_z} \;+\; \hat{h}(\{p\};\ \hat{I}_x, \hat{I}_y, \hat{I}_z), \qquad (1)$$

---

[1] It is well known that even a single nucleon may very profoundly disturb the rotational behavior of the whole nucleus through an alignment of its angular momentum with the temporary axis of rotation. In this study we consider the rotational behavior of 'pure' rotors i.e. uncoupled to individual nucleons. It will thus be of primary interest here to be able to connect the deviations from the 'standard' parabolic like behavior of the energy *vs.* spin relations to the deviations of the actual symmetry of a nucleus from the 'standard', ellipsoidal, $D_2$-symmetry.

where $\hat{h}$ contains terms that formally represent symmetries other than the $D_2$ symmetry and $\{p\}$ denotes the ensemble of Hamiltonian *parameters*.

## 2. Generalized Rotor Hamiltonian

In order to be able to conveniently represent various possible point-group symmetries we are going to introduce the basis of the tensor-operators

$$\hat{T}_{\lambda\mu}(n; \lambda_2, \lambda_3, \ldots \lambda_{n-1}) \equiv \left[\left(((\hat{I}\otimes\hat{I})_{\lambda_2}\otimes\hat{I})_{\lambda_3}\otimes \ldots \otimes\hat{I}\right)_{\lambda_{n-1}}\right]_{\lambda\mu} \quad (2)$$

where e.g. symbol $(\hat{I}\otimes\hat{I})_{\lambda_2}$ represents an ensemble of all components of the irreducible spherical tensors of rank $\lambda_2 = 0, 1$ or $2$, that are obtained through the Clebsch-Gordan coupling i.e.

$$(\hat{I}\otimes\hat{I})_{\lambda_2} \equiv \{ (\hat{I}\otimes\hat{I})_{\lambda_2\mu_2}; \quad \mu_2 = -\lambda_2, -\lambda_2 + 1, \ldots \lambda_2 \}. \quad (3)$$

The most important particular case of Eq. (2) corresponds to what we call *maximum stretching* situation where $\lambda_n = n$. In such a case we may simplify the notation without much ambiguity

$$\hat{T}_{\lambda\mu}(n; \lambda_2, \lambda_3, \ldots \lambda_{n-1}) \to \hat{T}_{\lambda\mu}(n). \quad (4)$$

Each of the above objects represents a spherical tensor operator of rank $\lambda$, an element of the basis constructed with the help of the rotor angular momentum operator $\{\hat{I}_{-1}, \hat{I}_0, \hat{I}_{+1}\}$. More precisely $\hat{T}_{\lambda\mu}(n)$ is a uniform polynomial of order $n$ with the coefficients defined through the Clebsch-Gordan coupling, where additionally $\lambda = n$. Since in general i.e. for the non-stretched couplings one has $\lambda \neq n$, we prefer to stress this fact in the notation below in which we keep explicitly $\lambda$ and $n$. In what follows we will parameterize the last expression in Eq. (1) in terms of sums of the $n^{th}$ order uniform polynomials that are at the same time tensors of rank $n$

$$\hat{h}(n, \lambda) \equiv \sum_{\mu=-\lambda}^{\lambda} c^*_{\lambda\mu}(n) \hat{T}_{\lambda\mu}(n), \quad (5)$$

where $c_{\lambda\mu}$ are arbitrary constants[2] i.e. they are $\vec{I}$-independent objects; as usual, they may, however, depend on any scalar function of the quantum number $I$ and this freedom will be used in Eq. (7) later on.

After all these preliminaries we may complete the definition of the generalized rotor Hamiltonian with

$$\hat{h}(\{p\}; \hat{I}_x, \hat{I}_y, \hat{I}_z) = \sum_{n=3}^{n_{max}} \hat{h}(n, \lambda), \quad (6)$$

---
[2] The only limitation imposed is that the resulting Hamiltonian remains hermitian.

where $\{p\}$ represents the ensemble of all the Hamiltonian parameters i.e. all the constants $\{c_{\lambda\mu}(n);\ \lambda = 3, 4, \ldots \lambda_{max}\}$.

## 3. Rotors and Symmetries

A strong motivation for studying the symmetries of the quantum rotating objects via the properties of the corresponding excitation spectra is provided by numerous examples in molecular physics (the reader is referred to the monograph [2] for a detailed discussion). There the symmetries provide several possible identification criteria, mainly through the characteristic degeneracies of excitation energies at a given spin. Often the degeneracies are satisfied only to a certain approximation (see below) and are experimentally manifested by a grouping of levels with certain quantum characteristics that are dictated by the considered symmetry.

In the recent *nuclear structure* literature the only remarkable series of discussions that deviate from the question of the $D_2$-symmetric rotors was the one related to the hypothetical $C_4$-symmetry in superdeformed (SD) nuclei, Ref. [3]. This idea originated from the experimental discovery of the so-called $(\Delta I = 2)$ - staggering $[(\Delta I = 4)$ - oscillations] in rotational bands of some SD nuclei. Although microscopic arguments in favor of such a mechanism in some superdeformed nuclei can be given in terms of $\alpha_{44}$-deformation driving orbitals, Ref. [4], the same reference finds no sufficient argument to justify the static $\alpha_{44}$-deformation hypothesis. No other microscopic calculations performed so far did confirm the possibility of an existence of *static* deformations of this kind in nuclei. The particle-rotor coupling mechanism that could increase the deviations from the regular energy *vs.* spin behavior cannot, after a model calculation of Ref. [5], replace the effect of a static deformation with the four-fold symmetry. To the contrary, arguments can be given that the staggering as defined and observed so far in experiment can be caused by a weak-interaction band-crossing.

Yet, the reviving discussion of a possibility that nuclei with symmetries non-trivially different from the $D_2$-symmetry exist in nature is to our opinion an important step forward in studying the potential richness of nuclear behavior. Indeed, theoretical predictions exist of *low-lying* isomeric states in nuclei that possibly manifest an approximate cubic symmetry ($T_d$-group symmetry) or a symmetry of a triangle ($C_3$ or $D_3$ symmetries), Refs. [6] and [7]. These symmetries correspond to the predicted (nearly) pure octupole shapes related to the deformations of the type $\alpha_{32}$ and $\alpha_{33}$. Thus the important message from the quoted references is that in many nuclei from the vicinity of the doubly-magic shell-closure nuclei there are large *octupole* type deformations that may develop. More precisely: they could either set in the ground-states before the quadrupole deformation eventually overtakes or they could set in the octupole deformation dominated isomers that compete energetically with the quadrupole deformation dominated ground-state minima [6].

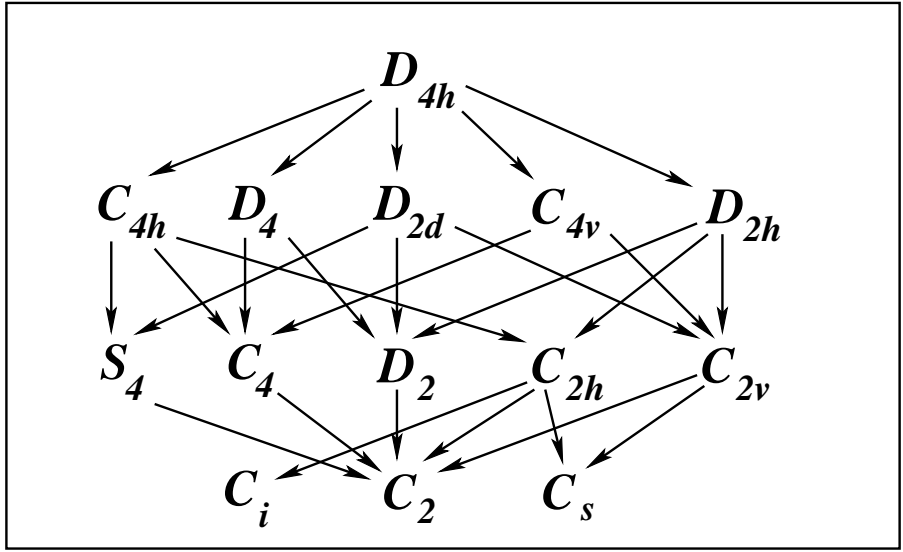

Fig. 1. Diagram of subgroups of the $D_{4h}$-group. It shows several possible nuclear symmetries that one needs to consider when the only criterion in mind is an existence of a 4-fold symmetry axis among the symmetries of the nucleus in question. These groups are: $D_{4h}$, $C_{4h}$, $D_4$, $D_{2d}$, $C_{4v}$, $S_4$ and $C_4$; in the literatures there has only been the latter possibility considered.

It will be one of our goals here to examine the quantum behavior of the rotors that are characterized by this type of exotic shapes. Before proceeding with that, however, we would like to discuss a number of questions that in the present context often have lead to confusion in the past. Here it will be convenient to use the $C_4$-symmetry case as a relatively recent one. When trying to explain the energy fluctuations in function of spin that produce 'oscillations' with the 'period' of $\Delta I = 4$ an existence of the 4-fold symmetry axis is a necessary (although far from sufficient) condition. In order to satisfy this condition a $C_4$-symmetry case has been considered. Figure 1 below shows most of the possibilities that should be *a priori* under consideration.

Although some of these symmetry groups may represent similar physical properties there are certainly several possibilities that remain unexplored.

Another aspect to consider is the physical meaning of the parameters that enter into the generalized rotor Hamiltonian of Eqs. (1) - (5). Although *a priori* all choices of these parameters are mathematically allowed as long as the resulting model Hamiltonian remains hermitian, in our study we would like to keep track of the role of the rotor moments of inertia $\mathcal{J}_x$, $\mathcal{J}_y$ and $\mathcal{J}_z$ in (1) as leading-role parameters that determine the dominating part of the Hamiltonian. In other words, when working with the well deformed nuclei we expect that the second term in (1) is 'small' (see below). This is a natural way to achieve the result that in nuclear physics of well deformed nuclei the deviations from the parabolic $E_I \sim I(I+1)$ rule are small. In our realization of the quantum rotors in this paper we therefore limit the liberty of the choice of the $c_{\lambda\mu}$ parameters and use

$$c_{\lambda\mu} = \frac{C_{\lambda\mu}}{[I(I+1)]^{(\lambda-1)/2}} \tag{7}$$

where $I$ is the angular momentum *quantum number* to be distinguished from the angular momentum *operator* $\hat{I}$ and $C_{\lambda\mu}$ are merely numerical constants. This is our way of defining the 'smallness' of the perturbing term with respect to what we consider as dominating: the traditional quadratic rotor Hamiltonian.

## 4. Advantages of a Tensor Representation of the Rotor Hamiltonian

In their excellent study of the barrier-penetration properties in the spin motion of the quantum rotors the authors of [3] have chosen an explicit-

$$\hat{H}_{H-M} = A\hat{I}_z^2 + B_1(\hat{I}_x^2 - \hat{I}_y^2)^2 + B_2(\hat{I}_x^2 + \hat{I}_y^2)^2 \tag{8}$$

rather than covariant-form (cf. Eqs. (5-6)) of the Hamiltonian definition; the former was well suited for the purposes of the reference quoted, the latter has several important advantages when discussing direct physical applications to nuclei as it will be illustrated below.

The tensor representation introduced earlier provides a basis in the mathematical sense and allows to express the fourth order terms in (8) as

$$(\hat{I}_x^2 - \hat{I}_y^2)^2 = \frac{3 - 4I^2}{5\sqrt{3}}\hat{T}_{00}(2) + \Big(\frac{3-4I^2}{\sqrt{150}} + \sqrt{\frac{3}{2}} + \frac{2}{35}\sqrt{6}(2-I^2)\Big)\hat{T}_{20}(2)$$
$$+ \sqrt{\frac{2}{35}}\hat{T}_{40}(4) + \hat{T}_{44}(4) + \hat{T}_{4-4}(4) \tag{9}$$

and

$$(\hat{I}_x^2 + \hat{I}_y^2)^2 = \frac{1}{105}\Big(-\frac{140\sqrt{3}}{3}I^2 - \frac{7}{3}(4I^2 - 3)\Big)\hat{T}_{00}(2)$$

$$-\left(175\sqrt{\frac{2}{3}}I^2 + 7\sqrt{\frac{2}{3}}(3-4I^2) + 9\sqrt{6}(2-I^2)\right)\hat{T}_{20}(2)$$
$$+ 6\sqrt{70}\,\hat{T}_{40}(4). \tag{10}$$

From the above expressions it becomes clear that the apparently fourth-order form of the Hamiltonian in Eq. (8) is in fact composed of a generic fourth-order (tensor) expressions that remain in that order when performing any orthogonal transformations of this Hamiltonian in space and of the second and zeroth order (tensor) operator expressions preceded by the quadratic (scalar) functions of the spin quantum number $I$. One can easily show that the 'usual' quadratic rotor Hamiltonian (i.e. the one with the constant coefficients) can be expressed as

$$\frac{\hat{I}_x^2}{2\mathcal{J}_x} + \frac{\hat{I}_y^2}{2\mathcal{J}_y} + \frac{\hat{I}_z^2}{2\mathcal{J}_z} = b_{00}\,\hat{T}_{00}(2) + b_{20}\,\hat{T}_{20}(2) + b_{22}[\hat{T}_{22}(2) + \hat{T}_{2-2}(2)] \tag{11}$$

where $b_{\alpha\beta}$ are some numerical coefficients - known functions of $\{\mathcal{J}_x, \mathcal{J}_y, \mathcal{J}_z\}$.

Consider an 'academic' case of a $C_4$-symmetric ($\mathcal{J}_x = \mathcal{J}_y$) Hamiltonian:

$$\hat{\mathcal{H}} = \frac{\hat{I}_x^2}{2\mathcal{J}_x} + \frac{\hat{I}_y^2}{2\mathcal{J}_y} + \frac{\hat{I}_z^2}{2\mathcal{J}_z} + B_1(\hat{I}_x^2 - \hat{I}_y^2)^2 + B_2(\hat{I}_x^2 + \hat{I}_y^2)^2. \tag{12}$$

Using tensor expressions of Eqs. (9)-(10) we may show an instructive identity

$$\hat{\mathcal{H}} = \frac{\hat{I}_x^2}{2\tilde{\mathcal{J}}_x(I)} + \frac{\hat{I}_y^2}{2\tilde{\mathcal{J}}_y(I)} + \frac{\hat{I}_z^2}{2\tilde{\mathcal{J}}_z(I)}$$
$$+ \left(\sqrt{\frac{2}{35}} + 6\sqrt{70}\right)\hat{T}_{40}(4) + \hat{T}_{4+4}(4) + \hat{T}_{4-4}(4). \tag{13}$$

The latter expression shows that the $C_4$-symmetric operator (8) is a sum of a $C_4$-symmetric tensor operator of the 4-th order and of a quadratic rotor Hamiltonian whose coefficients (moments of inertia $\{\tilde{\mathcal{J}}_x(I), \tilde{\mathcal{J}}_y(I), \tilde{\mathcal{J}}_z(I)\}$) are strongly varying with spin. Indeed, these coefficients can be explicitly calculated for any set of constants $B_1$ and $B_2$ and the corresponding results are illustrated in Fig. 2, at $B_2 = 0$, as an example.

Most of the nuclei known so far do *not* produce any evidence for the $C_4$-symmetry and thus we may believe that, should occasionally such a phenomenon arise, this could only be because of a perturbation of the usual $D_2$-type rotor structure by a $C_4$-type *admixture*. In such a case one would choose the moments of inertia $\{\mathcal{J}_i;\ i=x,y,z\}$ in Eqs.(1), (11)-(12), that correspond roughly to the expected rotational band curvature in accordance with experiment and then increase the effect of the fourth-order term (possibly modifying the $\mathcal{J}_i$ parameters *slightly*) to obtain at the same time the staggering and the right order of magnitude of (nearly constant) moments of inertia. Such an intuitive approach does not seem possible with Hamiltonian (8) as the tensor expansion demonstrates and Fig. 3 illustrates. More generally, an advantage of using tensor expansions in (1), is the possibility of taking into account the experimental fact that most of the nuclei seem to produce the $D_2$-rotor type behavior with a possible *slight* modifications due to other symmetry admixtures. Such a separation comes naturally within the tensor-operator description.

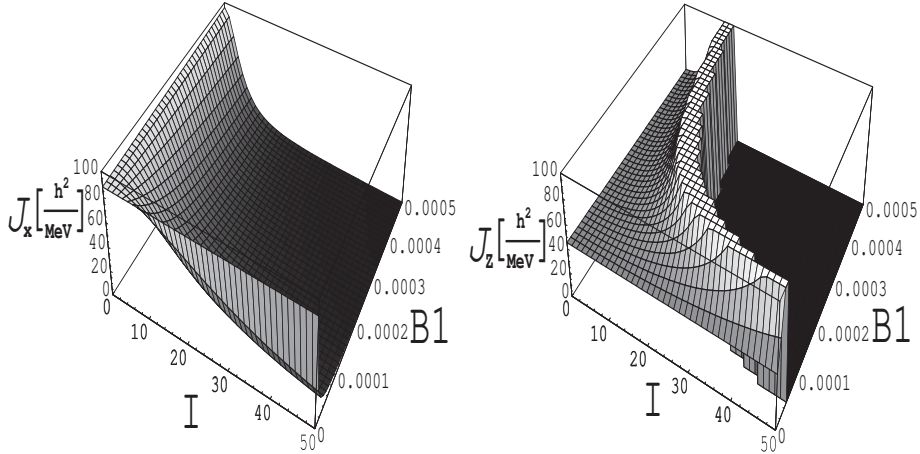

Fig. 2. The $C_4$-symmetric hamiltonian of Ref.[3] can be represented as a sum of the traditional-looking second order Hamiltonian with the spin-dependent moments of inertia and the fourth order $C_4$-symmetric tensor term. The figure illustrates the $\tilde{\mathcal{J}}_x(I)$ (left) and $\tilde{\mathcal{J}}_z(I)$ (right), in function of $I$ for various choices of the $B_1$ parameter. Parameter $B_2 = 0$ here; the staggering is implied in a part of the $(I, B_1)$-plane.

## 5. Octupole-Deformed Nuclei and the Corresponding Quantum Rotors

A different situation takes place if the nuclei with small quadrupole deformations are considered where the higher order multipoles contribute importantly or even dominate in the nuclear shape description. Calculations of Refs. [6] and [7] suggest that in the transitional nuclei with $Z$- and $N$-numbers slightly in excess of the doubly-magic shell closures (similarly for doubly-magic nuclei with a few proton and/or neutron holes) such situations may take place either in the ground-states or in the low-lying isomeric minima. The corresponding symmetries correspond to $\lambda = 3$ in Eqs. (1)

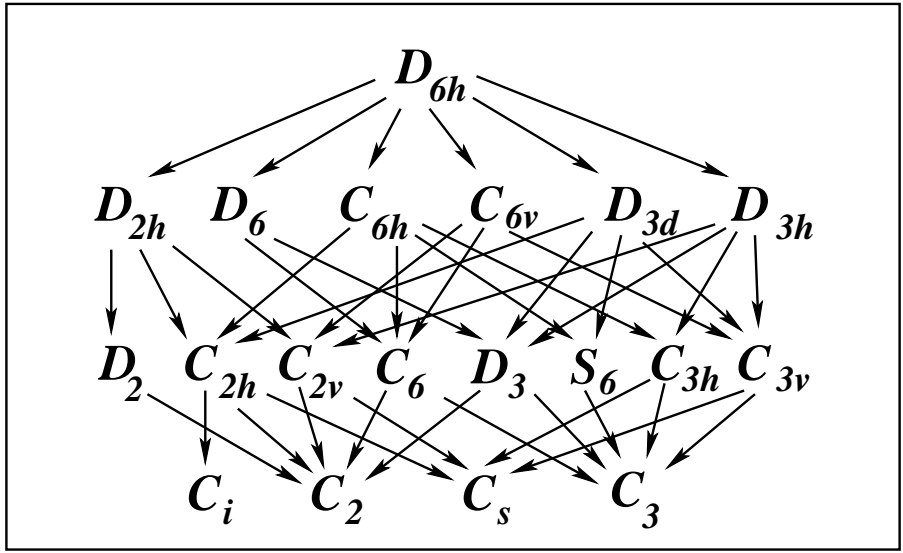

Fig. 3. Diagram of subgroups of the $D_{6h}$-group. A three-fold symmetry axis appears as a 'secondary' symmetry element in all the groups containing a six-fold symmetry and we are not going to discuss these cases here. Instead let us point out that the other groups containing a three-fold axis are: $D_{3d}$, $D_{3h}$, $D_3$, $C_{3h}$, $C_{3v}$ and $C_3$.

and (5); they may contain, among others the mass-asymmetry degrees of freedom ($\lambda = 3, \mu = 0$) or alternatively a three-fold symmetry axis as one of the symmetry elements ($\lambda = 3, \mu = 3$). (Other symmetries that may arise when using higher order multipole operators will be discussed elsewhere and below we will limit ourselves to presenting only very few illustrations focused on the two cases just mentioned).

From the theory of symmetry point of view the three-fold axis may appear as an element of several point groups as illustrated in Fig. 3. There are therefore several forms of the quantum rotor Hamiltonians that may appear in the three-fold axis context; as mentioned we limit the discussion to the case of the $C_3$ group.

Let us begin with an axially-symmetric rotor whose Hamiltonian contains the $\{\lambda = 3, \ \mu = 0\}$ terms in its definition (1). The mathematical structure of this third-order term is modeling the symmetry of a pear-shape nuclear mass distribution of an otherwise complicated, rotating many-body nuclear system. In the following we would like to illustrate the evolution of the rotor spectra when the corresponding coupling constant $C_{3,0}$ increases. For an axially symmetric rotor the $K$ quantum number is conserved and the corresponding symmetry group is $C_\infty$. In the discussion it will be convenient to distinguish the states that have $K = 0$ from the other states that will be arbitrarily divided into two groups (distinguished by two shades of gray colors) that correspond by definition to positive-$K$ values (related states are labeled with [K1]), and to negative-$K$ values (labeled with [K2]). We follow here our earlier suggestion that the effective nuclear rotors can be represented by the leading quadratic term plus the third order modifying term (cf. Eq. (1)). Since the microscopic calculations suggest the existence of such states for rather small quadrupole deformations we set in the model calculations of the moments of inertia $\alpha_{2,0} = 0.1$, that for Z=86 and N=132 gives by using a uniform nucleonic density ansatz, $\mathcal{J}_x = \mathcal{J}_y = 13.3\hbar^2/MeV$ and $\mathcal{J}_z = 12.1\hbar^2/MeV$. Strictly speaking we take the proportion between the two numbers above as suggested by the geometrical model and reduce the absolute values by a common factor to simulate the difference between the rigid-like and pairing dominated nucleus. Figure 4 represents the spectrum obtained with $C_{3,0} = 0.001$ while Fig. 5 analogous results for $C_{3,0} = 0.01$.

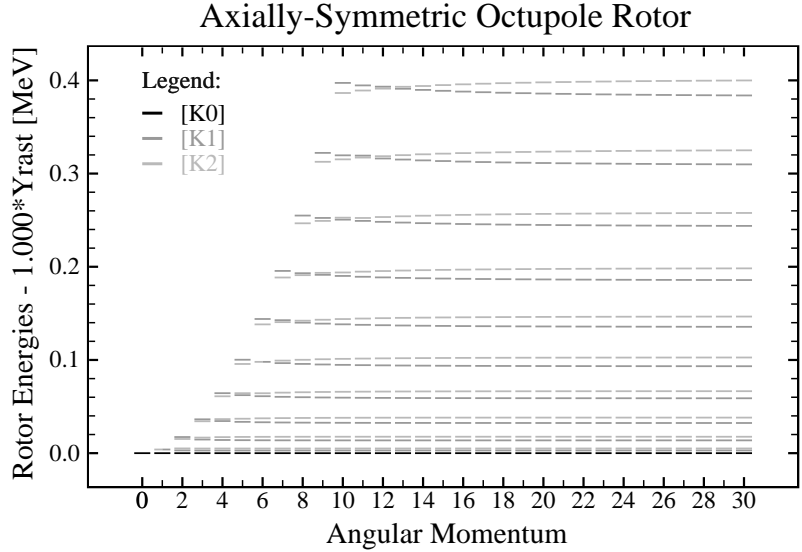

Fig. 4. Spectrum of an axially-symmetric rotor with the moments of inertia corresponding to the quadrupole deformation of $\alpha_{2,0} = 0.1$ (see text); $C_{3,0} = 0.001$.

Let us mention that at the present stage we have no way of determining the size of this constant in the Hamiltonian directly on the microscopic grounds; this is not so in the case of the moments of inertia of the ellipsoidal rotor. One can see that the excitation of states [K1] relative to the yrast line decreases with increasing spin while that of states [K2] increases, as the comparison of both Figures demonstrates. At the same time we can observe that increasing the coupling constant leads to a rearrangement of the spectrum; in particular the $K = 0$ sequence is not anymore the yrast and is getting further away from the yrast position.

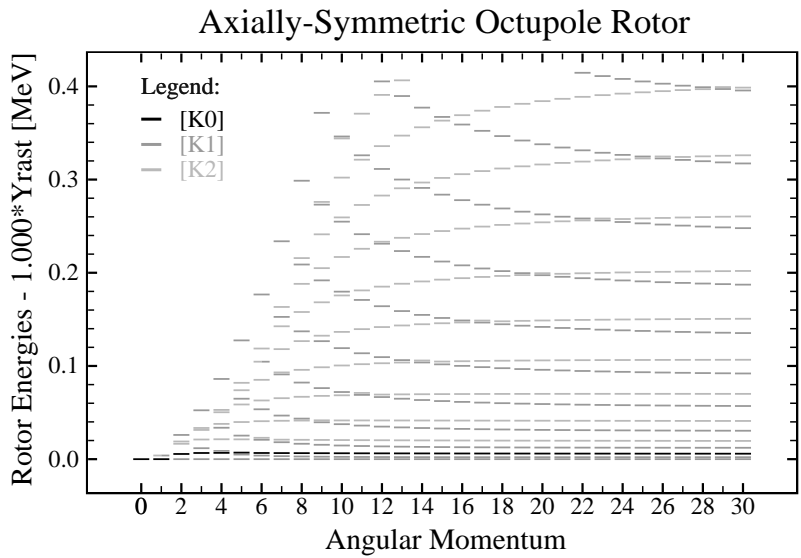

Fig. 5. Similar to Fig. 4 but including the 'octupole' term with $C_{3,0} = 0.01$; observe the progressive splitting of the positive-$K$ vs. negative-$K$ partner states.

There are several interesting properties of such 'octupole' rotors but we will have to limit ourselves to one illustration only that presents the expectation values of the angular momentum operators and the $(\Delta I = 2)$-staggering properties of the discussed *axially symmetric* octupole rotor, cf. Fig. 6. It is indeed interesting to observe a characteristic staggering in the illustrated observables that, when displayed in terms of the yrast energy, can reach several hundreds of keV in the model situation chosen here.

### 6. Quantum Rotors with Three-fold Symmetry Axis

In this section we are going to illustrate just a couple of spectroscopic features of a quantum rotor with a three-fold symmetry axis and a minor quadrupole (axially symmetric) deformation equal to that used in the previous case of an axial symmetry, cf. Figs. 7 and 8 and compare to Figs. 4 and 5. In the case of the $C_3$ group discussed here, there are three irreducible representations denoted [A1], [A2] and [A3], that generate three families of the rotational states. (The mathematical and interpretation aspects of the physics of related rotors will be discussed in more detail in a forthcoming publication). Observe that with a small-strength third order perturbative terms in the Hamiltonian the states are grouped characteristically: first there is the yrast line composed of *single* states of [A1] representation that is accompanied by *doubly* degenerate *excited* bands belonging to the same irreducible representation. These double degeneracies are only approximate; they vary from a fraction of $eV$ to several $keV$. In addition we have in between the $[A1]$ sequences two well spaced sequences of excited bands. Each of them is composed of doublets of nearly degenerated pairs of states; in this latter case, however, each member of a doublet belongs to a different irreducible representation, $[A2]$ or $[A2]$, and thus corresponds to a different symmetry. We can directly observe the role of the three-fold symmetry that manifests itself through the triplets of nearly degenerate states at the lower part of the spectrum. A discussion of many interesting features of the rotors of that kind will have to be postponed to the forthcoming publication.

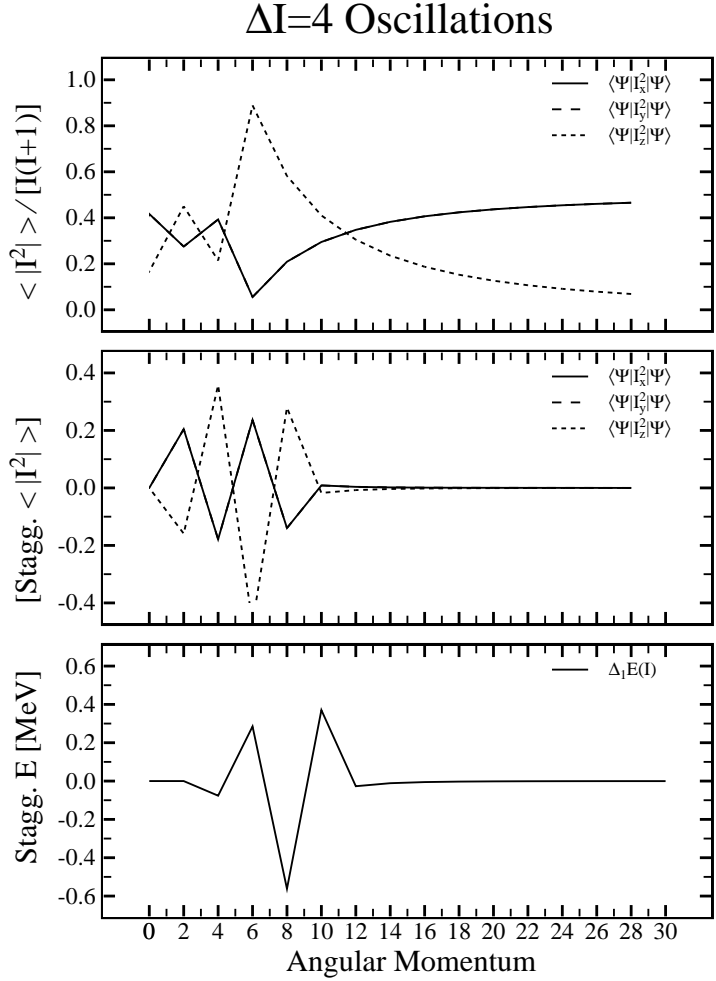

Fig. 6. Three-dimensional aspects of the motion of an axial 'octupole' rotor represented through the normalized expectation values of $\hat{I}_x^2$, $\hat{I}_y^2$ and $\hat{I}_z^2$, top, compared to the staggering properties of the same objects, middle part, and to the energy staggering, bottom, for the yrast line of the $C_{3,0} = 0.01$ case. (The staggering observable is defined as in Ref. [3]).

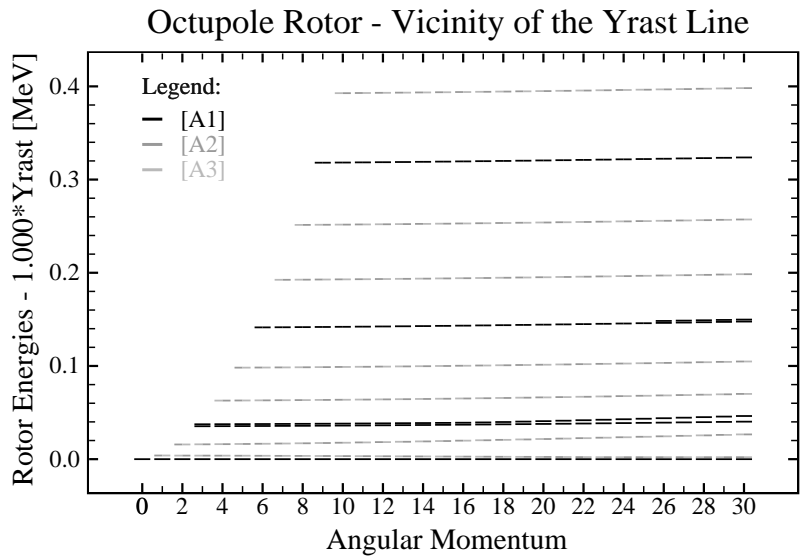

Fig. 7. The quantum rotor spectrum obtained with $C_{3,3} = 0.001$; the quadratic term in the rotor Hamiltonian has the moments of inertia that correspond to a small quadrupole deformation of $\beta_2 = 0.10$, see captions to Figs. 4 and 5.

## 7. Summary and Conclusions

In this article we examine the concept of the generalized-rotor Hamiltonians. They are built-up using spherical-tensor operators, the latter constructed with the help of angular momentum components $\{\hat{I}_{-1}, \hat{I}_0, \hat{I}_{+1}\}$. Using these tensor operators we arrive at the structure of the Hamiltonians that is well adapted to the nuclear physics applications i.e. a sum of a quadratic, often dominating term with an ellipsoidal (or spherical) symmetry and of higher order terms responsible for the (expected to be small) exotic-symmetry admixtures. We illustrate the functioning of this mathematical formulation using $C_4$-symmetric rotors discussed previously by other authors.

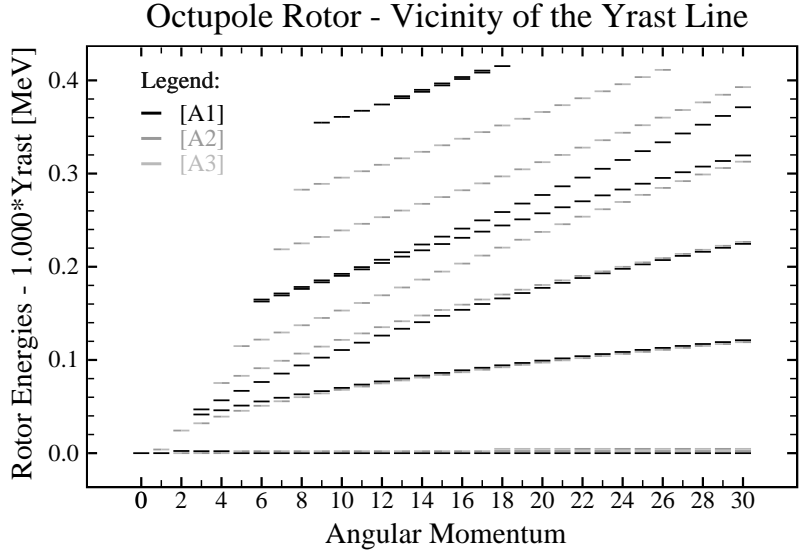

Fig. 8. Similar to that in Fig. 7, but for $C_{3,3} = 0.01$; observe that in the lower-right sector of the figure the spectrum is composed of nearly degenerate <u>triplets</u> of states characteristic of the presence of a three-fold symmetry axis among the Hamiltonian symmetries. In the upper sector of the Figure we find the states grouped into nearly degenerate doublets that were characteristic of a 'weak octupole coupling' of Fig. 7. In between the two areas there is a 'separatrix' region for which none of the two asymptotic behavior applies.

Next we follow up the results of the earlier microscopic calculations suggesting an existence of isomeric configurations with exotic symmetries. We have limited our illustrations to those dominated by components with $\{\lambda = 3, \mu = 3\}$, $C_3$-group, and $\{\lambda = 3, \mu = 0\}$, $C_\infty$-group. Both these symmetries induce the spectra that deviate characteristically from those of the well known ellipsoidal-rotors. We suggest the use of these differences for a possible identification of these symmetries through experiments.

An interesting question related to the symmetries is that of an experimental verification. At first glance the predicted spectroscopic features may seem easy to identify because of the characteristic dependence of the energy *vs.* spin. However in reality, there are several complications to be expected. First of all, the yrast and low lying excited bands of interest are expected to have very similar moments of inertia and consequently they are likely to remain unresolved. This aspect is certainly the one that will challenge the new class of gamma-ray tracking detection systems; we believe that the signals sought exist in many experimental data already taken. Secondly, the symmetry-manifesting behavior of the bands in question that are in a sense much more interesting than the other ones, will be disturbed by pairing and by individual nucleonic alignment. This latter aspect, is a question to theory to select the minima that are preferably free from 'back-bending' and that 'keep' the paring correlations at a more or less constant level.

A support from the France-Poland scientific exchange program *POLONIUM* and from the $IN_2P_3$, France, is acknowledged.